\documentclass[twocolumn,twocolappendix]{aastex701}
\usepackage{amssymb,amsmath,amsfonts,graphicx,epsf}
\usepackage{makecell}
\usepackage{soul}
\usepackage[normalem]{ulem}

\newcommand{\gtorder}{\mathrel{\raise.3ex\hbox{$>$}\mkern-14mu\lower0.6ex\hbox{$\sim$}}}
\newcommand{\ltorder}{\mathrel{\raise.3ex\hbox{$<$}\mkern-14mu\lower0.6ex\hbox{$\sim$}}}

\shorttitle{The Radius of PSR~J0614$-$3329}
\shortauthors{Miller, Dittmann, Holt, et al.}

\usepackage{hyperref}

\begin{document}

\title{THE RADIUS OF THE NEUTRON STAR PSR J0614$-$3329 FROM NICER DATA
}

\correspondingauthor{M.~C.~Miller}
\email{mcmiller@umd.edu}

\author[0000-0002-2666-728X]{M.~C.~Miller}
\affiliation{Department of Astronomy and Joint Space-Science Institute, University of Maryland, College Park, MD 20742-2421 USA}
\email{miller@astro.umd.edu}

\author[0000-0001-6157-6722]{A.~J.~Dittmann}
\altaffiliation{NASA Einstein Fellow}
\affiliation{Institute for Advanced Study, 1 Einstein Drive, Princeton, NJ 08540, USA}
\email{dittmann@ias.edu}

\author[0000-0002-3097-942X]{I.~M.~Holt}
\affiliation{Department of Astronomy and Joint Space-Science Institute, University of Maryland, College Park, MD 20742-2421 USA}
\email{imholt@umd.edu}

\author[0000-0002-3862-7402]{F.~K.~Lamb}
\affiliation{Illinois Center for Advanced Studies of the Universe and Department of Physics, University of Illinois at Urbana-Champaign, 1110 West Green Street, Urbana, IL 61801-3080, USA}
\affiliation{Department of Astronomy, University of Illinois at Urbana-Champaign, 1002 West Green Street, Urbana, IL 61801-3074, USA}
\email{fkl@illinois.edu}

\author[0000-0003-2759-1368]{C.~Chirenti}
\affiliation{Department of Astronomy, University of Maryland, College Park, MD 20742-2421, USA}
\affiliation{Astroparticle Physics Laboratory NASA/GSFC, Greenbelt, MD 20771, USA}
\affiliation{Center for Research and Exploration in Space Science and Technology, NASA/GSFC, Greenbelt, MD 20771, USA}
\email{chirenti@umd.edu}

\author[0009-0008-6187-8753]{Z.~ Arzoumanian}
\affiliation{X-Ray Astrophysics Laboratory, NASA Goddard Space Flight Center, Code 662, Greenbelt, MD 20771, USA}
\email{zaven.arzoumanian-1@nasa.gov}

\author[0000-0003-4962-145X]{J.~Berteaud}
\affiliation{Department of Physics, Royal Holloway, University of London, Egham, TW20 0EX, UK}
\email{joanna.berteaud@rhul.ac.uk}

\author[0000-0002-9870-2742]{S.~Bogdanov}
\affiliation{Columbia Astrophysics Laboratory, Columbia University, 550 West 120th Street, New York, NY 10027 USA}
\email{slavko@astro.columbia.edu}

\author[0000-0001-7115-2819]{K.~C.~Gendreau}
\affiliation{X-Ray Astrophysics Laboratory, NASA Goddard Space Flight Center, Code 662, Greenbelt, MD 20771, USA}
\email{keith.c.gendreau@nasa.gov}

\author[0000-0001-6119-859X]{A.~K.~Harding}
\affiliation{Theoretical Division, Los Alamos National Laboratory, Los Alamos, NM 87545, USA}
\email{ahardingx@yahoo.com}

\author[0000-0002-6089-6836]{W.~C.~G. Ho}
\affiliation{Department of Physics and Astronomy, Haverford College, 370 Lancaster Avenue, Haverford, PA 19041, USA}
\email{who@haverford.edu}

\author[0000-0003-1080-5286]{C.~Kalapotharakos}
\affiliation{Astrophysics Science Division, NASA Goddard Space Flight Center, Greenbelt, MD 20771, USA}
\email{constantinos.kalapotharakos@nasa.gov}

\author[0000-0003-4357-0575]{S.~M. Morsink}
\affiliation{Department of Physics, University of Alberta, Edmonton, AB, T6G 2E1, Canada}
\email{morsink@ualberta.ca}

\author[0000-0002-5297-5278]{P.~S. Ray}
\affiliation{Space Science Division, U.S. Naval Research Laboratory, Washington, DC 20375, USA}
\email{paul.s.ray3.civ@us.navy.mil}

\author[0000-0003-4815-0481]{R.~A.~Remillard}
\altaffiliation{Deceased 28 July 2026}
\affiliation{MIT Kavli Institute for Astrophysics \& Space Research, MIT, 70 Vassar Street, Cambridge, MA 02139, USA}
\email{ronrem4@gmail.com}

\author[0000-0002-9249-0515]{Z. Wadiasingh}
\affiliation{Department of Astronomy, University of Maryland, College Park, MD 20742-2421, USA}
\affiliation{Astrophysics Science Division, NASA Goddard Space Flight Center, Greenbelt, MD 20771, USA}
\affiliation{Center for Research and Exploration in Space Science and Technology, NASA/GSFC, Greenbelt, MD 20771, USA}
\email{zorawar.wadiasingh@nasa.gov}

\author[0000-0002-4013-5650]{M.~T.~Wolff}
\affiliation{Space Science Division, U.S. Naval Research Laboratory, Washington, DC 20375, USA}
\email{MTWolff@cox.net}

\begin{abstract}

Neutron star radius measurements, particularly using X-ray data collected with the Neutron star Interior Composition Explorer (NICER), have provided invaluable information for models of cold, catalyzed matter at densities above that of nuclear saturation.  Here we present an analysis of NICER data on the 318-Hz pulsar PSR~J0614$-$3329, which has a mass $\sim 1.4$--$1.5~M_\odot$ determined from radio observations.  The best-fitting model we explore has three uniform-temperature circular hot spots and yields a symmetric 68\% credible range for the equatorial circumferential radius of 9.88--12.77~km.  We also explore joint fits to the NICER data and the X-ray Multi-Mirror (XMM-Newton) data on this pulsar, but find that the models that best fit both data sets systematically underpredict the XMM-Newton data when, as is standard, it is assumed that the XMM-Newton background is known from observations of surrounding fields.  Finally, we discuss the implications of our results, combined with data on other neutron stars, for the properties of the dense matter in neutron star cores.

\end{abstract}


\section{INTRODUCTION}
\label{sec:introduction}

The properties of the matter in cores of neutron stars, which has densities above the nuclear saturation density $n_0\approx 0.16~{\rm fm}^{-3}$, is  much colder than its Fermi temperature, and is highly isospin asymmetric, cannot be explored in terrestrial laboratories or predicted from first principles.  Astronomical observations that constrain the properties of this cold, equilibrated, high-density matter are therefore crucial. During the last $\sim 15$ years, numerous data sets have contributed to our understanding of this matter. Analyses of data on X-ray pulsars acquired using NASA's Neutron star Interior Composition Explorer (NICER) (see \citealt{2016SPIE.9905E..1HG}) have played a particularly critical role.

NICER's key capability is that it can determine the arrival time of individual X-ray photons with a precision of $\sim 100$~nsec or better while also determining their energy to within $10\%$ or better. When folded at a pulsar's rotational period, the NICER counts in each energy channel form a waveform in pulse phase. The data sets we analyze are therefore two-dimensional, consisting of the number of counts at each phase in each energy channel. When properly analyzed, these data produce constraints on the mass and radius of the star and hence on the equation of state (EOS) of the matter in it. (In this paper ``mass" will always mean the star's gravitational mass and ``radius" will always mean the star's equatorial circumferential radius.)

NICER's neutron star targets are active radio pulsars and are therefore not accreting. This means that X-ray photons emitted by the stellar surface can propagate to the NICER detectors without interacting with accreting plasma in the system and that each photon emitted by the star can be mapped to a precise position on the star's surface.

The NICER mission was motivated in part by early studies using synthetic data \citep{2009ApJ...705L..36L,2009ApJ...706..417L,2013ApJ...776...19L,2015ApJ...808...31M,2016RvMP...88b1001W} which indicated that phase-resolved data, unlike phase-integrated data, could be analyzed in a way that is robust against many types of systematic errors (see \citealt{2010ApJ...722...33S,2013arXiv1312.0029M,2014MNRAS.445.4218K}, and \citealt{2016EPJA...52...63M} for caveats on radius estimates made using data from thermonuclear X-ray bursts or quiescent low-mass X-ray binaries, and \citealt{2017A&A...608A..31N} for a possible improvement based on non-uniform surface emission).  These studies have since been extended to include joint fits of NICER and X-ray Multi-Mirror Newton data \citep{2025arXiv251116759H}.  

Pulsars with previous NICER-based radius estimates include PSR~J0030$+$0451 \citep{2019ApJ...887L..24M,2019ApJ...887L..21R,2024ApJ...961...62V,2026ApJ..1005..201K}, PSR~J0740$+$6620 \citep{2021ApJ...918L..28M,2021ApJ...918L..27R,2022ApJ...941..150S,2024ApJ...974..295D,2024ApJ...974..294S}, PSR~J0437$-$4715 \citep{2024ApJ...971L..20C,2026ApJ..1000L..48M}, PSR~1231$-$1411 \citep{2024ApJ...976...58S,2025ApJ...981...99Q}, PSR~J0614$-$3329 \citep{2025ApJ...995...60M}, and PSR~J2124$-$3358 \citep{2026arXiv260703721G}.

Here we present a new analysis of NICER data on PSR~J0614$-$3329.  In Section~\ref{sec:observations} we describe our extraction of the data, in Section~\ref{sec:methods} we provide a brief overview of our methods, and in Section~\ref{sec:results} we present our results.  In Section~\ref{sec:EOS} we discuss the implications of our results for the properties of cold, equilibrated matter at densities above the density of nuclear saturation.  We summarize our conclusions in Section~\ref{sec:conclusions}.  

In Appendix~\ref{sec:XMM} we present results from joint analyses of NICER and XMM-Newton data, including a comparison with the results of \citet{2025ApJ...995...60M} and show that when we make the standard assumption that we know the XMM-Newton background, the joint fit systematically underpredicts the observed XMM-Newton counts.  
Appendix~\ref{sec:apppost} provides
further details of our results, including plots of the posteriors and corner plots.
Posterior samples from our analysis are available on Zenodo via \href{https://doi.org/10.5281/zenodo.22131748}{doi:10.5281/zenodo.22131748}.

\section{OBSERVATIONS}
\label{sec:observations}

\subsection{NICER XTI data}
\label{sec:NICERdata}
The NICER X-ray Timing Instrument (XTI) data for PSR J0614$-$3329 used in this analysis were collected between 2018 March 3 and 2023 May 15, corresponding to ObsID range 0030050101--6030050328, with a total unfiltered exposure of 1.6 Ms. The event lists from the individual ObsIDs were subjected to initial screening using the \texttt{nicerl2} task in the NICERDAS v12 software that is part of HEASoft version 6.33.2, along with the instrument calibration file  \texttt{xti20240206}.  Additional event filtering was carried out using the \texttt{psrpipe} script from  NICERsoft\footnote{Available at \url{https://github.com/paulray/NICERsoft}.} with similar criteria to previous analyses. Time intervals were removed when any of the following events occurred: 1)~excessive optical loading, as indicated by undershoot rates exceeding 200 counts s$^{-1}$; 2)~high particle background exceeding 1.5 counts s$^{-1}$ per Focal Plane Module; 3)~elevated solar activity, indicated by a Potsdam planetary geomagnetic index  $K_p\ge5$; 4)~space station passages through regions with a low cut-off rigidity, namely, \texttt{COR\_SAX}~$\le1.5$~GeV/c. A subsequent cut was made by excising any 16\,s intervals with a 2--10~keV countrate above 1.0~counts~ s$^{-1}$, yielding 1.09~Ms of effective exposure time.

Redistribution matrix (RMF) and effective area (ARF) files tailored to the final set of good-time intervals (GTIs) were constructed using the \texttt{nicerl3} task in NICERDAS. Pulse phases were assigned to the final event list using the \texttt{photonphase} script in \texttt{PINT} \citep{2021ApJ...911...45L} with a timing ephemeris obtained using Nan\c{c}ay Radio Telescope and Fermi Large Area Telescope data that covered the time span of the NICER observations (M. Kerr, personal communication).

In the analyses we discuss in this paper we used NICER pulse invariant (PI) channels 30--199 inclusive.

\subsection{XMM-Newton EPIC data}
\label{sec:XMMdata}
We used an archival XMM-Newton European Photon Imaging Camera (EPIC) observation of PSR~J0614$-$3329 acquired on 2010 October 4. The MOS1/2 and pn instruments were operated in Full Window mode, which does not offer sufficient time resolution to study the 
neutron star's pulsations.
Starting with the unprocessed ODF data products, we ran the event files through the \texttt{emproc} and \texttt{epproc} pipelines in SAS version 22.1.0. Next, we screened for instances of elevated background flaring by removing any 100-second segments where the count rate over the entire detector at $\ge$10~keV exceeded 0.3~counts~s$^{-1}$ for MOS1/2 and 0.4~counts~s$^{-1}$ for pn. This data processing and filtering procedure resulted in final effective exposure times of 14.4 ks for MOS1/2 and 11.3 ks for pn. The recommended event quality parameter cuts were also made, with \texttt{FLAG}\,=\,0 and $\texttt{PATTERN}\,\le12$ for MOS1/2 and  $\texttt{PATTERN}\,\le4$ for pn.  

The final source spectra were extracted from circular regions of radius 32$''$ centered on the radio pulsar position. 
Background estimates were obtained using source-free regions near the target. For MOS1/2 a circular region of radius 125$''$ on the central CCD at detector coordinates $X=28515.278$, $Y=21561.578$ was chosen whereas for pn a 60$''$ circular region on the same CCD as the source at detector coordinates $X=22892.089$, $Y=30026.528$ was used.

In the analyses we discuss in Sections~\ref{sec:joint} and \ref{sec:comparison}, we used data from XMM-pn PI channels 57--399 inclusive, from XMM-MOS1 PI channels 60--399 inclusive, and from XMM-MOS2 PI channels 60--399 inclusive.

\section{OVERVIEW OF METHODS}
\label{sec:methods}

Our approach to modeling NICER and, where relevant, XMM-Newton data has been summarized in several previous papers \citep{2019ApJ...887L..25B,2019ApJ...887L..26B,2019ApJ...887L..24M,2021ApJ...914L..15B,2021ApJ...918L..28M,2022ApJ...941..150S,2024ApJ...974..295D,2024ApJ...975..202C,2026ApJ..1000L..48M}, which contain additional details, cross-code comparisons, and justifications.  Other groups have used the same basic approach but with different ray-tracing codes, statistical samplers, and assumptions about the background \citep{2019ApJ...887L..21R,2021ApJ...918L..27R,2024ApJ...971L..20C,2024ApJ...974..294S,2024ApJ...961...62V,2025ApJ...995...60M,2025ApJ...981...99Q,2026ApJ..1005..201K}.  

In brief, our basic assumptions are as follows:

\begin{enumerate}

\item  Some regions of the neutron star's surface are bombarded by highly relativistic particles, which have been accelerated within the star's magnetosphere toward the star's surface. This flux of relativistic particles deposits energy deep beneath the star's surface and this energy heats the surface in those regions where it is being bombarded. As a result, these regions emit X-rays.

\item  The locally measured spectrum and angular dependence of the X-ray emission from each element on the stellar surface depends on the temperature of the element, the chemical composition of the star's atmosphere, and the local strength of the stellar magnetic field. To model this emission we use the emission from a fully ionized nonmagnetic hydrogen atmosphere computed by \citet{2001MNRAS.327.1081H}.

\item We model the pattern of heated regions on the stellar surface phenomenologically, using uniform-temperature circular or oval regions (``hot spots"). These hot spots have arbitrary locations and sizes and, in the case of oval spots, arbitrary aspect ratios and orientations. We number the spots, which are allowed to overlap. If a surface element is at a position that is covered by more than one spot, we assume it emits with the effective temperature of the lowest-numbered spot.

\item For ray tracing from the star to the observer we use the ``oblate Schwarzschild approximation" (\citealt{2007ApJ...663.1244M,2014ApJ...791...78A,2025ApJ...994..163J}), in which the surface has, to a good approximation, the correct oblate shape produced by the star's rotation but the actual external spacetime is approximated by the Schwarzschild spacetime.

\item We assume that counts in the NICER data that do not come from hot spots are not modulated at a frequency commensurate with the rotation of the neutron star and thus treat these counts by including a parameter for each NICER PI channel that represents the background counts in that channel and is independent of the star's rotational phase and any background counts in other PI channels.  This treatment is valid even if the background changes with time; the only assumption is that the background counts populate rotational phases uniformly. Ultimately, we
marginalize over these background 
parameters,
independently for each NICER PI channel.  Unlike in PSR~J0437$-$4715 (see \citealt{2026ApJ..1000L..48M}), there is no evidence in the PSR~J0614$-$3329 NICER or XMM-Newton data for a nonthermal component modulated at the star's rotational frequency.

\end{enumerate}

As in our previous analyses of pulsar waveforms, our inference is Bayesian, in that we explicitly specify the models and priors that we assume.  As in \citet{2026ApJ..1000L..48M}, our primary sampler is \texttt{pocoMC} \citep{2022MNRAS.516.1644K}.   See Sections 3.7 and 3.8 of \citet{2026ApJ..1000L..48M} for more details about our statistical samplers and Figure~12 of \citet{2026ApJ..1000L..48M} for a comparison of the radius posterior for PSR~J0437$-$4715 from \texttt{pocoMC} with the very similar radius posterior from the final sampling with \texttt{emcee}.

\section{RESULTS OF OUR ANALYSIS OF NICER PULSE WAVEFORM DATA}
\label{sec:results}

We fit three classes of surface-emission models to the NICER data: two circular spots, two oval spots, and three circular spots.
All models were run with progressively higher \texttt{pocoMC} resolution until the evidence converged. For our headline three-circle model, we performed additional sampling using \texttt{emcee} to refine the posterior.  Table~\ref{tab:wf-primary-parameters} lists our primary model parameters and their priors. The \texttt{emcee} posteriors samples from our three-circle model can be found \href{https://zenodo.org/records/22131748}{on Zenodo}.

\begin{deluxetable*}{c|l|c}
\setlength{\tabcolsep}{19pt} 
    \tablecaption{Primary parameters of the pulse waveform models considered in this work.}
\tablehead{
      \colhead{Parameter} & \colhead{Definition} & \colhead{Assumed Prior}
    \label{tab:wf-primary-parameters}
}
\startdata
      \hline
      $c^2R_e/(GM)$ & Inverse of stellar compactness & $3.2-8.0$ \\
      \hline
      $M/M_\odot$ & Gravitational mass & ${\cal N}(1.44,0.07)$\\
      \hline
      $\theta_{\rm c}$ (rad) & Colatitude of spot center & $0 - \pi$ \\
      \hline
      $\Delta\theta$ (rad) & Spot half-extension& $0-3$ \\
      \hline
      $kT_{\rm eff}$ (keV) & Spot effective temperature & $0.011-0.5$\\
      \hline
      $f$ & Spot elongation factor &\makecell{ $\log_{10}f$ flat from $0$ to $+1$} \\
      \hline
      $\psi$ (rad) & Spot tilt angle & $0-\pi$\\
      \hline
      $\Delta\phi$ (cycles) & Spot longitude difference & $0-1$ \\
      \hline
      $\theta_{\rm obs}$ (rad) & Observer inclination & ${\cal N}(1.5411,0.00424)$ \\
      \hline
      $N_H$ (cm$^{-2})$ & Neutral H column density & $(0-20)\times 10^{20}$ \\
      \hline
      $d$ (kpc) & Distance & ${\cal N}(0.585,0.045)$ \\
      \hline
      $f_{\rm NICER,eff}$ & NICER effective area factor & ${\cal N}(1,0.104)$\\
\enddata
\tablecomments{Main parameters and priors, where the priors on mass and observer inclination come from \citet{2025MNRAS.536.1467M} and the distance prior comes from \citet{2016MNRAS.455.3806B}.
The spot parameters describe an oval spot; circular spots have elongation factor 1 and tilt angle 0.  The spot parameters are selected independently from each other.  By definition the longitude of the center of spot 1 is 0, and the spot longitudes of other spots are with respect to spot 1 (note that after the spots have been constructed, the overall phase of the waveform is marginalized with respect to the data). }
\end{deluxetable*}

\subsection{Model comparison}
\label{sec:fitquality}

\begin{deluxetable*}{c|l|c}
    \tablecaption{Summary of the results of the analyses using our pulse waveform models.}
\tablewidth{0pt}
\tablehead{
      \colhead{Model} & \colhead{$\Delta\ln{\cal Z}$} &  \colhead{$\pm 1\sigma$ $R_{\rm eq}$ (km)}
    \label{tab:NICERonly}
}
\startdata
      \hline
      2 circles &  $-10.42$ & $12.11-13.76$ \\
      \hline 
      2 ovals & $-11.47$ & $11.63-15.56$ \\
      \hline
      \textbf{3 circles} & \textbf{0} & \textbf{9.88~--~12.77}\\
      \hline
\enddata
\tablecomments{Log evidence relative to our favored model (in boldface), and $\pm 1\sigma$ posterior radius range, for our three models.}
\end{deluxetable*}

Table~\ref{tab:NICERonly} summarizes the Bayesian evidence for each model, together with the symmetric (about the median) 68\% credible interval of the inferred radius.   We see that the radius ranges overlap among the three models, but that the 3-circle model is clearly preferred, by $\Delta\ln{\cal Z}>10$.   

\subsection{Characteristics of the preferred model with three circular spots}

\begin{figure}
          \includegraphics[width=\linewidth]{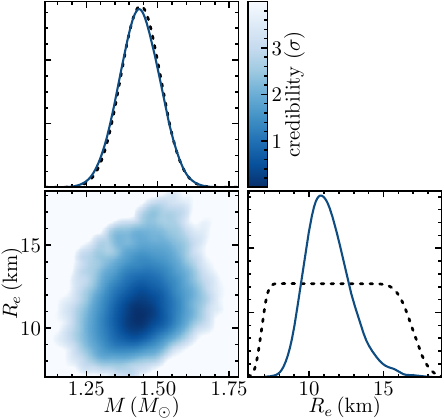}
\caption{Mass and radius posteriors (solid lines in the one-dimensional plots; color weighting in the two-dimensional plot), along with their one-dimensional priors (dashed lines), for our featured general three-circular-spot model fit to the NICER data.  The mass posterior is clearly determined almost entirely by the prior, but the radius posterior is strongly informed by the X-ray data.}
\label{fig:mr}
\end{figure}

\begin{figure}
\includegraphics[width=\linewidth]{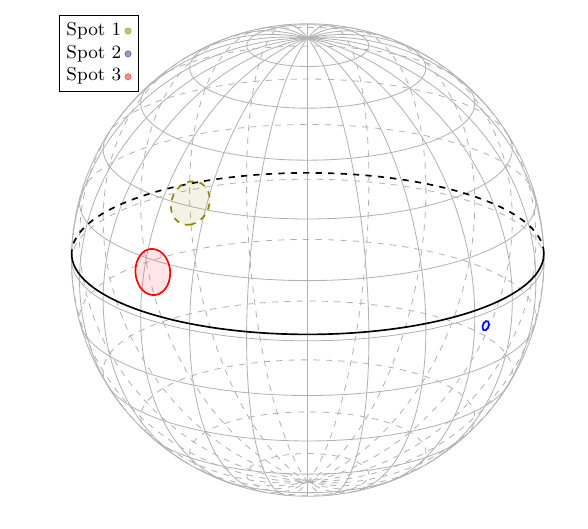}
\caption{
The hot-spot geometry in our 
maximum-likelihood 3-circle fit to the NICER data. The black line marks the inferred 
(dominated by the prior from pulse timing)
line-of-sight inclination at which we view the pulsar. Solid lines depict parts of the stellar surface with surface normal vectors pointing out of the page, whereas dashed lines are used to plot parts of the surface with surface normal vectors pointing into the page. The spot colors are simply qualitative labels.  Other spot configurations also fit the data well (see Appendix~\ref{sec:apppost} for some examples), including ones with a large spot ($\Delta\theta>1$ radian), but large-spot configurations are disfavored by $\Delta\ln{\cal L}>6$ compared with the combination shown here. Spot 1 is colored olive; spot 2, the hottest, is colored blue; and spot 3, the coolest, is colored red. }
\label{fig:spots}
\end{figure}

\begin{figure}
          \includegraphics[width=\linewidth]{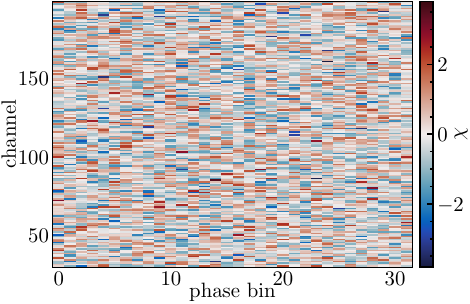}
\caption{Phase-channel residuals to NICER data from our maximum-likelihood three-circle fit to those data.  The color indicates the value of $\chi=(d-m)/m^{1/2}$ for data $d$ and model $m$.  NICER PI channel increases upward, whereas the rotational phase bin increases to the right.  The lack of any pattern in the residuals, along with the good total $\chi^2/{\rm dof}=5154.89/5252$ indicate that the fit is adequate based on this test.}
\label{fig:residuals}
\end{figure}

\begin{figure}          \includegraphics[width=\linewidth]{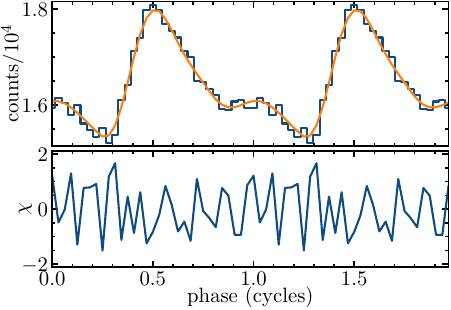}
\caption{Top panel: bolometric counts from the NICER data (histogram) and the NICER predictions from the best-fitting three-circle model (orange curve).  Bottom panel: $\chi=(d-m)/m^{1/2}$ for the bolometric data, where $d$ is the number of observed NICER counts and $m$ is the predicted number for each of the 32 rotational phase bins.  Again, no patterns are evident and there are no notable outliers, suggesting that the model passes this test of goodness of fit.}    
\label{fig:boloresiduals}
\end{figure}

Figure~\ref{fig:mr} shows the mass-radius posterior in our preferred three-circle fit to the NICER data, along with marginalizations
of this posterior into mass and radius separately.  The dashed lines in the one-dimensional plots show the priors.  The mass posterior is essentially the same as the prior, but the radius posterior is strongly informed by the X-ray data.

Figure~\ref{fig:spots} displays the best-fitting parameter combination from our 3-circle model to the NICER data.  
Note that other spot configurations also provide good fits to the data; in particular, there are some configurations with a single much larger spot ($\Delta\theta>1$ radian).  However, large-spot configurations  are strongly disfavored  compared with the spot pattern shown in Figure~\ref{fig:spots}.

Non-antipodal emitting-region geometries are favored throughout most of the posterior probability. This is consistent with a surface magnetic field configuration more complex than a centered dipole, possibly including the presence of multipolar components (see, e.g., \citealt{2019ApJ...887L..23B,2020ApJ...893L..38C,2021ApJ...907...63K}).  Physically self-consistent multiwavelength modeling, in which X-ray, $\gamma$-ray, and radio observables constrain a common magnetic-field and viewing geometry, would provide complementary information on the geometry.

\subsection{Goodness of fit checks}

The maximum-likelihood 3-circle fit to the NICER data has a phase-channel $\chi^2/{\rm dof}=5154.89/5252$.  The probability of this $\chi^2$ or larger for a correct model with this number of degrees of freedom is 83\%.  The bolometric $\chi^2/{\rm dof}=26.42/27$ (probability 50\%).  Neither probability is low enough to raise concerns, so although this does not prove that it is the right model, it passes these tests.  See Appendix~\ref{sec:apppost} for the full posteriors for our fits with three circular spots to the NICER data. 

Figure~\ref{fig:residuals} shows the phase-channel residuals $\chi$ for the maximum-likelihood 3-circle model, defined as
\begin{equation}
    \chi={\rm (data-model)}/({\rm model})^{1/2}\; ;
\label{eq:chi}
\end{equation}
note that the data and model are both represented in counts, so $\chi$ is dimensionless.  Similarly, Figure~\ref{fig:boloresiduals} shows the bolometric (i.e., channel-summed) version of Figure~\ref{fig:residuals}.  No patterns or notable outliers are evident in either residuals plot.

\section{IMPLICATIONS FOR THE EOS}
\label{sec:EOS}

The majority of the properties of neutron stars 
that we can measure depend only on the equation of state (EOS) of the matter in them. (A notable exception is neutron star cooling rates, which can be used to obtain information about the microscopic properties of the matter in their interiors; see \citealt{2025AN....34640110H} for a recent review.)  The EOS of a given material is the pressure in the material as a function of intensive quantities, such as the density, temperature, and composition.  The thermal temperatures of the matter in the interiors of the neutron stars observed using NICER are thought to be $\ltorder 10^{8-9}$~K, much lower than the Fermi temperatures $T_F$ of the matter in their cores, which are $\sim 10^{12}$~K. Hence the thermal pressure in the interiors of these neutron stars does not significantly affect their macroscopic properties.  

It is also expected that the matter in the cores of the neutron stars observed using NICER, where the density is comparable to or greater than the saturation density of nuclear matter, is in chemical equilibrium.  Therefore the pressure $P$ in this region can be expressed as a function of a single quantity,
such as
the energy density $\epsilon$ or the baryon number density $n$, and this equation of state $P(\epsilon)$ or $P(n)$ is a universal function for  cold, equilibrated matter at densities above the density of nuclear saturation.
Thus, although neutron stars with different masses will have different densities in their cores, it is expected that at a given $\epsilon$ or $n$ the pressure will be the same in any neutron star.

If a neutron star is not rotating, its central density and the EOS of the matter in it determine its mass and structure. Hence, once we know the EOS of neutron-star matter, we can compute the structure of a nonrotating star with a given central density using the relevant structure equations. In particular, we can determine the maximum mass of a nonrotating star composed of matter with this EOS. We can also compute the radius, tidal deformability, moment of inertia, and other macroscopic properties of a star of any mass. Rotation alters the star's structure, but for the neutron stars that have been observed using NICER, all of which have rotation frequencies $\nu_{\rm rot}<400$~Hz, the effect of rotation on these properties is much smaller than the measurement uncertainties (see, e.g., Figure~13 of \citealt{2019ApJ...887L..24M}).  

Our procedure for inferring the EOS of neutron-star matter follows that of \citet{2020ApJ...888...12M}.  We start with a set of EOSs, which we assume all have the same prior probability.  For a given EOS, we calculate the likelihood of each of the independent observational data sets, given that EOS, and then multiply their likelihoods to obtain an overall likelihood of the data given that EOS.  We then repeat this procedure for each EOS. This gives the posterior probability distribution for the trial EOSs.  Using this new, reweighted distribution of EOSs, we derive quantities such as the posterior probability distribution of the maximum mass, the radius at a given mass, or anything else that is predicted by the EOS.  

In more detail, we assume that we know the EOS from zero density to some threshold density, and then use a framework to extrapolate to higher densities.  We assume that we know the EOS to half of nuclear saturation density, because this is approximately where the crust transitions to the core \citep{2013ApJ...773...11H}.  In previous work we used the QHC19 EOS \citep{2019ApJ...885...42B} up to half of saturation density.  In contrast, here our EOS below half of staturation density is the more microscopically realistic QHC21 EOS \citep{2022ApJ...934...46K}, which was inspired in part by NICER analyses of the $\sim 2.1~M_\odot$ pulsar PSR~J0740$+$6620 and which uses the chiral effective field theory EOS \citep{2020PhRvR...2b2033L,2021PhRvC.103d5808D,2021ARNPS..71..403D} below 1.5 times saturation density.  The QHC19 and QHC21 low-density priors give nearly indistinguishable EOS posteriors after neutron star observations are included. 

We then need to decide on our EOS extrapolation framework.  Because of its flexibility, and for continuity with previous papers, we use the Gaussian processes framework described in \citet{2021ApJ...918L..28M}, and in fact use the same $10^5$ equations of state above $n_s/2$ as in previous papers \citep{2021ApJ...918L..28M,2024ApJ...974..295D,2026ApJ..1000L..48M} so that any differences come from added data (and as a minor contribution, the assumptions about the EOS below $n_s/2$) rather than from the EOS framework.

The final decision to make is the data sets to include in our analyses.  As in \citet{2026ApJ..1000L..48M} we use the updated masses of PSR~J1614$-$2230 ($M=1.937\pm 0.014~M_\odot$; see \citealt{2023ApJ...951L...9A}) and PSR~J0348$+$0432 ($M=1.806\pm 0.037~M_\odot$; see \citealt{2025ApJ...983L..20S}); the tidal deformability from the double neutron star merger GW170817 \citep{2017PhRvL.119p1101A} and the probable double neutron star merger GW190425 \citep{2020ApJ...892L...3A}; and the NICER-based (sometimes including XMM-Newton data) mass and radius estimates of PSR~J0030$+$0451 \citep{2019ApJ...887L..24M}, PSR~J0740$+$6620 \citep{2024ApJ...974..295D}, and PSR~J0437$-$4715 \citep{2026ApJ..1000L..48M}, to which we add the measurement of PSR~J0614$-$3329 in the current paper.  

For the mass-radius and mass-deformability data sets we used kernel density estimation with a Gaussian kernel that has 0.1 times the bandwidth recommended by \citet{silverman1986}; see Section~5.1 of \citet{2021ApJ...918L..28M} for details.  We also included a constraint on the nuclear symmetry energy at saturation density (defined as the energy per nucleon of pure neutron matter minus the energy per nucleon of matter with an equal number of neutrons and protons) of $S=32\pm 2$~MeV \citep{2012PhRvC..86a5803T}, where we imposed beta equilibrium including muons using the approach in Section~II of \citet{2016arXiv160408575B}.  Finally, we explored the consequences of including the high estimated mass of $M=2.35\pm 0.11~M_\odot$ from PSR~J0952$-$0607 (\citealt{2026ApJ...996..101R}; note that this pulsar has a high rotation rate of 707~Hz and that \citealt{2026ApJ...996..101R} find that the zero-rotation mass equivalent is $M=2.32\pm 0.11~M_\odot$, which is what we use).  As emphasized by \citet{2026ApJ...996..101R}, although the best-estimate mass for this pulsar is high and therefore constraining, concerns about systematics in the modeling of PSR~J0952$-$0607 and similar ``spider" pulsars suggests caution in using them in EOS constraints.

\begin{deluxetable}{ccccc}
\setlength{\tabcolsep}{2.3pt}
\caption{Maximum Masses and Fiducial Radii Including the PSR~J0614$-$3329 Measurements Reported Here}
\tablehead{
\colhead{Quantity} & \colhead{Data set} &  \colhead{$-1\sigma$} & \colhead{median} & \colhead{$+1\sigma$}
}         
\startdata
$M_{\rm TOV}(M_\odot)$&\citet{2026ApJ..1000L..48M}&2.07&2.20&2.44 \\
$M_{\rm TOV}(M_\odot)$&This work&2.02 &2.15 &2.38 \\
$M_{\rm TOV}(M_\odot)$&PSR~J0952$-$0607&2.26 &2.41 &2.59 \\
\hline
$R_e(1.4~M_\odot)({\rm km})$&\citet{2026ApJ..1000L..48M}&12.09 &12.56 &13.03 \\
$R_e(1.4~M_\odot)({\rm km})$&This work&11.91 &12.44 &12.88 \\
$R_e(1.4~M_\odot)({\rm km})$&PSR~J0952$-$0607&11.88 &12.38 &12.82 \\
\enddata
\tablecomments{
Median and $\pm1\sigma$
values of the maximum mass $M_{\rm TOV}$ of a non-rotating neutron star (where TOV refers to the papers \citealt{1939PhRv...55..364T} and \citealt{1939PhRv...55..374O}) and the radius $R_e(1.4~M_\odot)$ of a neutron star with a fiducial mass $1.4~M_\odot$, from our analysis prior to our PSR~J0614$-$3329 measurement, then including that measurement, and then including the mass measurement of PSR~J0952$-$0607 by \citet{2026ApJ...996..101R}, spin-corrected to $2.32\pm 0.11~M_\odot$ (see text for details). We see that the addition of our comparatively small measured radius for PSR~J0614$-$3329 decreases the estimated maximum mass and fiducial radius slightly, and that further inclusion of the estimated mass of PSR~J0952$-$0607 increases the quantiles for the maximum mass substantially but only lowers the quantiles for the fiducial radius by a small amount.}
\label{tab:maxmass}
\end{deluxetable}

\begin{figure}
          \includegraphics[width=\linewidth]{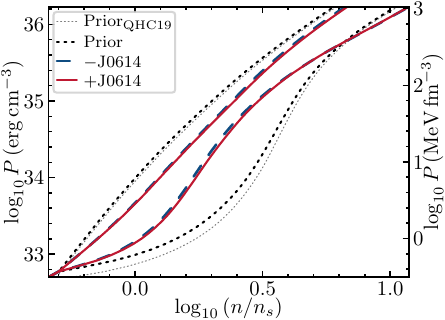}
\caption{Posterior range in log pressure as a function of log baryon number density in units of the saturation density $n_s=0.16~{\rm fm}^{-3}$, for our QHC21 prior (black dotted line; the grey dotted line shows the QHC19 prior), our posterior using the QHC21 prior with all information other than our PSR~J0614$-$3329 measurement (blue dashed line), and our posterior with all information including our PSR~J0614$-$3329 measurement (red solid line).  For each line type, at a given number density the lower curve shows the 5th percentile of the pressure, and the upper curve shows the 95th percentile of the pressure.  We do not include constraints based on the mass of PSR~J0952$-$0607 from \citet{2026ApJ...996..101R}.  The posteriors (not shown) that use the QHC19 low-density prior are essentially indistinguishable from the posteriors that we display here.  We see that our results for PSR~J0614$-$3329 are broadly consistent with our previous analyses, as indicated by the small change that inclusion of this measurement makes to the pressure posterior.  See the main text for additional details.}
\label{fig:EOS}
\end{figure}

Given these assumptions and data sets, the main results are in Table~\ref{tab:maxmass} (which shows the symmetric 68\% credible ranges for the maximum stable mass of a nonrotating neutron star and the radius of a fiducial $1.4~M_\odot$ neutron star) and Figure~\ref{fig:EOS}, which shows the symmetric 90\% credible range of the pressure as a function of baryon number density, in both cases emphasizing the effect of the new measurement.

\section{CONCLUSIONS}
\label{sec:conclusions}

Our analysis of NICER data on the $\sim 1.4~M_\odot$ pulsar PSR~J0614$-$3329 yields an estimate of its radius consistent with previous NICER-based radius measurements of the $\sim 1.4~M_\odot$ pulsars PSR~J0030 and PSR~J0437, though centered at a smaller radius.  The constraints we have obtained here on the EOS of matter at a few times nuclear saturation density, when combined with the constraints obtained from other analyses of astronomical data, including previous analyses of NICER data, continue to refine our understanding of the EOS of neutron-star matter, thereby further improving our understanding of the properties of cold, equilibrated matter at very high densities.

\section*{Acknowledgments}

The authors thank Gordon Baym and Toru Kojo for illuminating discussions regarding the equation of state of dense matter.  A.J.D. and M.C.M. were supported in part by NASA ADAP grant No. 80NSSC21K0649.  M.C.M. and C.C. were also supported in part by NASA grants 80NSSC25K7108, 80NSSC25K0511, and 80NSSC26K0667.  Part of this work was performed at the Aspen Center for Physics, which is supported by National Science Foundation grant PHY-2210452.  We are grateful to the Institute for Nuclear Theory at the University of Washington for its kind hospitality and stimulating research environment; this research was therefore supported in part by the INT's U.S. Department of Energy grant No.~DE-FG02-00ER41132. A.J.D. was supported by NASA through the Hubble Fellowship Program grant No.~HST-HF2-51553.001, awarded by the Space Telescope Science Institute, which is operated by the Association of Universities for Research in Astronomy, Inc., for NASA, under contract NAS5-26555. J.B. was supported by UKRI FLF grant number MR/Y018257/1 (PI: Graber). S.B. was supported in part by NASA grants 80NSSC20K0275, 80NSSC22K0728, and 80NSSC25K7544.  The research of S.M.M. is supported by NSERC Discovery Grant RGPIN-2026-05707. C.C. and Z.W. were supported by NASA under award number 80GSFC24M0006. C.K. was partly supported by NASA grants 22-ADAP22-0142 and 22-TCAN22-0027.  Some of the resources used in this work were provided by the NASA High-End Computing (HEC) Program through the NASA Center for Climate Simulation (NCCS) at Goddard Space Flight Center.  The authors acknowledge the University of Maryland supercomputing resources (http://hpcc.umd.edu) that were made available for conducting the research reported in this paper. Portions of this work performed at NRL were supported by NASA.  This research has made use of data products and software provided by the High Energy Astrophysics Science Archive Research Center (HEASARC), which is a service of the Astrophysics Science Division at NASA/GSFC and the High Energy Astrophysics Division of the Smithsonian Astrophysical Observatory.  We acknowledge extensive use of NASA's Astrophysics Data System (ADS) bibliographic services and arXiv.

\facility{\textit{NICER} (\citealt{2016SPIE.9905E..1HG})}

\software{emcee (\citealt{2013PASP..125..306F}), Python (\citealt{2007CSE.....9c..10O}), NumPy (\citealt{2020NumPy-Array}), Matplotlib (\citealt{2007CSE.....9...90H}), PGF/Ti$k$Z \citep{tantau:2013a}, Cython (\citealt{2011CSE....13b..31B}), schwimmbad (\citealt{schwimmbad}), HEASoft \citep{2014ascl.soft08004N}, PINT \citep{2021ApJ...911...45L}, and pocoMC (\citealt{2022MNRAS.516.1644K})}

\pagebreak

\appendix

\section{Joint Fits with XMM-Newton data}
\label{sec:XMM}

\subsection{Joint analyses}
\label{sec:joint}

Our approach to joint fits of NICER and XMM-Newton data is described in detail in \citet{2021ApJ...918L..28M} and \citet{2024ApJ...974..295D}.  To the parameters listed in Table~\ref{tab:wf-primary-parameters} we add effective area factors for the XMM-pn, XMM-MOS1, and XMM-MOS2 data.  For comparison with \citet{2025ApJ...995...60M}, our priors for these parameters are all independent normal distributions with means of 1 and standard deviations of 0.104 (i.e., the same as our prior on the NICER effective area factor), although in reality there are likely to be much tighter correlations between the effective area parameters of the XMM-Newton instruments.

 Because the time resolution and effective area of XMM-Newton are insufficient to yield useful phase-resolved data for PSR~J0614$-$3329, we need to treat the XMM-Newton background in a different way from the NICER background.  Our standard assumption, which is consistent with the assumption used by all groups in published joint NICER-XMM analyses, is that the only non-spot contribution to the XMM-Newton data is the measured background around PSR~J0614$-$3329 (see Section~\ref{sec:XMMdata}).  We further assume that the counts measured in the background represent Poisson samples from the expected background.  Thus if in some XMM-Newton PI channel $i$ there are $b_i$ measured background counts (where $b_i$ is an integer), the prior probability that the average background during that observation is $a_i$ is given by $q(a_i|b_i)=\left[a_i^{b_i}/b_i!\right]e^{-a_i}$.  Using this prior probability distribution for the background, and adjusting for the relative extraction areas on the sky and instrument responses for the source and background
regions, we marginalize over the background after multiplying by the likelihood of the XMM-Newton data given the spot model and the background.  

\begin{figure}
          \includegraphics[width=\linewidth]{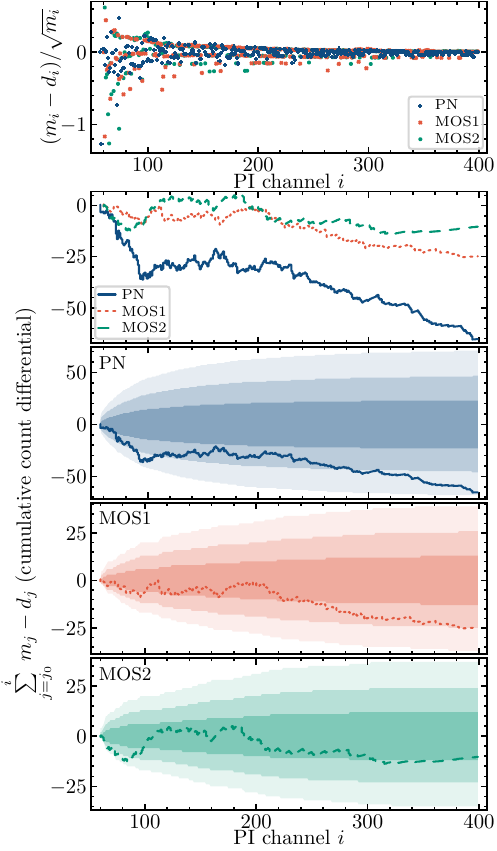}
\caption{
Comparison between the counts predicted by the best joint-fit model and the data for the three XMM-Newton instruments, as function of PI channel. The top panel displays the normalized residual in each PI channel. The lower panels display the cumulative sum, starting at the minimum PI channel for each instrument $j_0$ ($j_0=57$ for XMM-pn and $j_0=60$ for XMM-MOS1 and XMM-MOS2), of the difference between the model and data.
The second panel from the top shows the cumulative differences for XMM-pn, XMM-MOS1, and XMM-MOS2. The remaining three panels plot the cumulative differences for each instrument separately, combined with the $1\sigma$ (darkest region), $2\sigma$, and $3\sigma$ (lightest region) ranges expected from Poisson statistics if the model were correct.  We see that in all three instruments, most notably for XMM-pn, the predictions in the best joint-fit model fall short of the data.
}
\label{fig:cumXMM}
\end{figure}

However, when we use this assumption there is tension between the NICER and the XMM-Newton data, in the sense that the best joint fit underestimates the XMM-Newton counts.  This is evident in Figure~\ref{fig:cumXMM}, which shows the cumulative 
difference between the best joint-fit model and the data in each of the XMM-Newton instruments.  The XMM-MOS2, XMM-MOS1, and XMM-pn models are, respectively, $\approx 1\sigma$, $\approx 2\sigma$, and $\approx 3\sigma$ low compared with the data.  The NICER-only fits have an even greater deficit relative to the XMM-Newton data: for example, our maximum-likelihood NICER-only three-circle model has an XMM-pn deficit of 208.934 counts relative to the data, an XMM-MOS1 deficit of 72.059 counts relative to the data, and an XMM-MOS2 deficit of 53.263 counts relative to the data.

Although Figure~\ref{fig:cumXMM} only shows the results for the best fit to the joint NICER+XMM data sets, we have performed the same exercise for a set of 100 samples drawn from the posterior. Out of these, no model overpredicted the counts in any XMM-Newton instrument: the total model XMM-pn counts ranged from 444 to 500, versus 553 observed counts; the total model XMM-MOS1 counts ranged from 129 to 160, versus 165 observed counts; and the total model XMM-MOS2 counts ranged from 120 to 147, versus 148 observed counts.  We caution that these posterior samples were chosen by finding combinations with high NICER+XMM likelihoods  rather than reproducing the total counts in a given instrument.  Nonetheless, the systematic underestimation of the XMM-Newton data could suggest that there is additional background in the XMM-Newton data beyond what is captured in fields
near PSR~J0614$-$3329.  

The consequence is that in NICER-XMM joint fits without additional background, the fit assigns more of the unpulsed emission to the spots, which requires a more compact star to increase gravitational lensing and thus (given the well-constrained mass) requires a smaller radius.  This is why, using standard assumptions about XMM-Newton backgrounds, the inferred $+1\sigma$ radius drops from $12.765$~km in our NICER-only analysis to 11.88~km when XMM-Newton data are incorporated under the assumption of known background.

\begin{deluxetable*}{c|c}
    \tablecaption{Comparison of three-circle NICER+XMM joint fit results using only the known XMM-Newton background versus allowing for additional background as in \citet{2025arXiv251116759H}.  We also show our headline NICER-only result for comparison.}
\tablewidth{0pt}
\tablehead{
      \colhead{Background} & \colhead{$\pm 1\sigma$ $R_{\rm eq}$ (km)}
    \label{tab:XMMback}
}
\startdata
      \hline
      Standard & $9.52-11.88$ \\
      \hline 
      Additional & $11.60-14.76$ \\
      NICER-only & $9.88-12.77$\\
      \hline
\enddata
\end{deluxetable*}

We could instead use the approach of \citet{2025arXiv251116759H}, in which the XMM-Newton background is allowed to include an additional component.  In Table~\ref{tab:XMMback} we compare the resulting $\pm 1\sigma$ radius ranges between our standard background assumption (i.e., no additional background) and the case treated by \citet{2025arXiv251116759H}, which allows for additional background up to $\sim 4\times$ the known background.  We see that the assumption about the XMM-Newton background makes a large difference to the inferred radius.

Thus our standard background assumption leads to a discrepancy with the data, and the addition of unknown background changes the radius posterior significantly.  For these reasons, our headline result comes from our analysis of just the NICER data.

\subsection{Comparison with \citet{2025ApJ...995...60M}}
\label{sec:comparison}

The analysis of PSR~J0614$-$3329 by \citet{2025ApJ...995...60M} was a joint analysis of NICER and XMM-Newton observations.  Their headline result was a symmetric 68\% credible range in the radius of $10.29^{+1.01}_{-0.86}$~km.  This result used the ``ST+PDT" model in the publicly available X-PSI package \citep{2023zndo...7632629R}.  In this model, one spot (ST, for single temperature) is a uniform-temperature circle, whereas the other (PDT, for protruding double temperature) is two uniform-temperature circles on top of each other with at least a point of overlap, and pixels in the overlap region emit with the effective temperature of the ``superseding" circle.  No parts of the ST and PDT spots can overlap each other, and all three circles must be less than or equal to $\pi/2$ in angular radius.  Thus ST+PDT is a subset of our three-circle model.

\citet{2025ApJ...995...60M} use the same NICER data that we do, but an older extraction of the XMM-Newton data.  Thus for our comparison tests, we use the same XMM-Newton source and background data as in \citet{2025ApJ...995...60M}.  

\citet{2025ApJ...995...60M}  use the same approach to the NICER background that we do, but treat the XMM-Newton background differently.  In our analysis as well as that of \citet{2025ApJ...995...60M}, the background data are counts in the region around the PSR~J0614$-$3329 pointing, and take the form of counts in the PI channels of each of the three XMM-Newton instruments.  Both groups estimate the probability distribution of the long-term average number of counts based on the observed counts, and both groups then apply a scaling factor between the background data and the PSR~J0614$-$3329 pointing.  The difference lies in the translation between observed background counts and the probability distribution of the long-term average number of background counts.

Suppose that in a given XMM-Newton PI channel $i$, there are $b_i$ counts in the background observation.  The task is to determine the prior distribution of the expected number $a_i$, that is, to determine $q(a_i|b_i)$.  Then the two approaches are: 

\begin{enumerate}

\item As we described in Section~\ref{sec:joint}, we assume that the counts are drawn from a Poisson distribution, so that $q(a_i|b_i)=\left[a_i^{b_i}/b_i!\right]e^{-a_i}$.

\item \citet{2025ApJ...995...60M} follow \citet{2021ApJ...918L..27R} in their approach.  The prior on $a_i$ is flat between $b_i-3\sqrt{b_i}$ and $b_i+3\sqrt{b_i}$.  If $b_i>0$ but $b_i-3\sqrt{b_i}<0$ then the prior is flat between $0$ and $b_i+3\sqrt{b_i}$.  If $b_i=0$ then $b_i$ is set to the next nonzero background count number going upward in PI channel and the prior on $a_i$ is determined using this new $b_i$.

\end{enumerate}

The prior from either approach is then multiplied by the likelihood of the data, in channel $i$, in the pointed observation given the assumed background and the counts contributed by the spot model under consideration.  This posterior is then marginalized over the background to get the factor contributed by that channel in that data set for the given spot model.

To compare our results with those of \citet{2025ApJ...995...60M} we fit the data using the ST+PDT model with the same priors as in \citet{2025ApJ...995...60M} and use their treatment of the XMM-Newton background.  Using \texttt{pocoMC}, with repeated runs with higher precision until apparent convergence, we find $R_{\rm eq}=10.30^{+1.37}_{-1.02}$~km, compared with $R_{\rm eq}=10.29^{+1.01}_{-0.86}$~km from \citet{2025ApJ...995...60M}.  Thus for the same data, model, and background treatment our symmetric
68\% credible region is 28\% broader than that of \citet{2025ApJ...995...60M}.  Based on previous comparisons \citep{2016S&C....26..383B,2020AJ....159...73N,2023StSur..17..169B,2023MNRAS.521.1184L,2024OJAp....7E..79D}, we suspect that this is because the \texttt{MultiNest} precision parameters used by \citet{2025ApJ...995...60M} (in this case, $2\times 10^4$ live points and a sampling efficiency of 0.05) are insufficient for convergence.

\section{Posterior Distributions}
\label{sec:apppost}

\begin{figure*}
    \includegraphics[width=\linewidth]{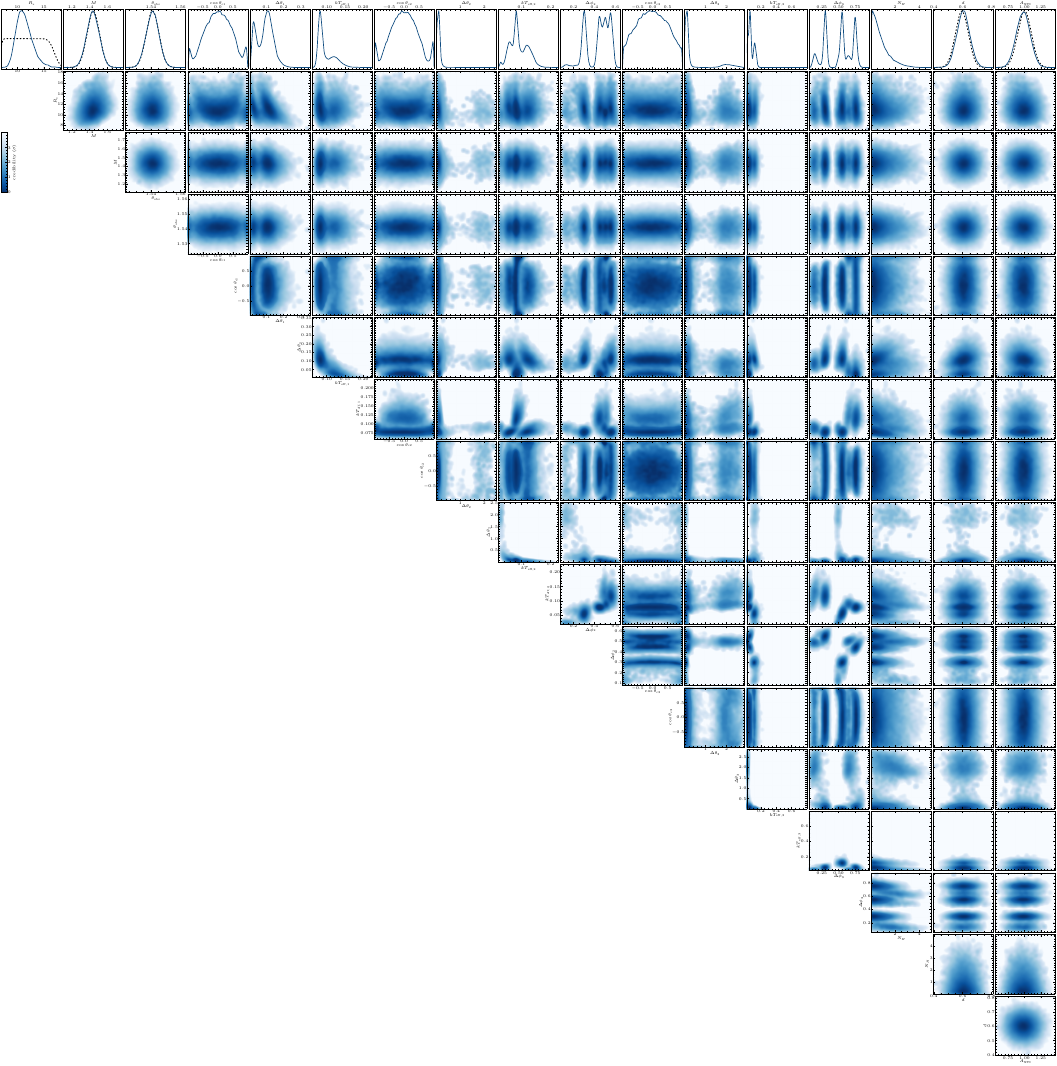}
    \caption{Posterior probability density distributions from our analysis of the NICER data on PSR~J0614$-$3329, where the units are the same as in Table~\ref{tab:posteriors}.  This corner plot 
displays results for
    our featured model, in which we have three uniform-temperature circular spots with arbitrary overlap. 
    Dotted traces in the one-dimensional distributions depict prior information.
    }
    \label{fig:posteriors}    
\end{figure*}

\begin{figure*}
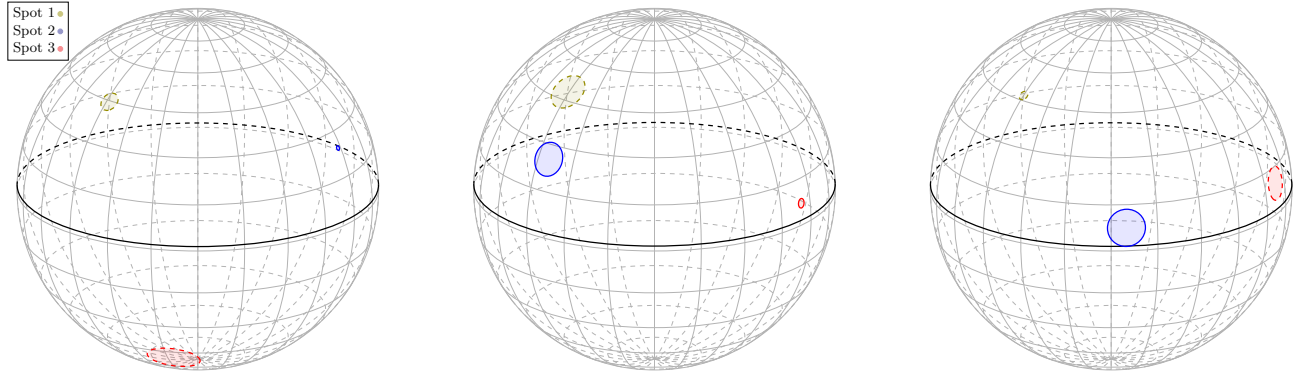

\gridline{\fig{{j0614_3circ_paper-mode1}.pdf}{0.33\textwidth}{}
          \fig{{j0614_3circ_paper-mode3}.pdf}{0.33\textwidth}{}
          \fig{{j0614_3circ_paper-mode4}.pdf}{0.33\textwidth}{}}
    \caption{Best-fitting spot configurations within different modes (see Figure \ref{fig:posteriors}), where we have excluded the mode in which the global best fit resides (see Figure \ref{fig:spots}) and ignored the subdominant modes that include very large spots. In each case the first, second, and third spots are colored olive, blue, and red respectively. As before, solid lines are used to plot parts of the stellar surface with    surface normal vectors pointing out of the page, while dashed lines are used to plot parts of the surface with surface normal vectors pointing into the page. }\label{fig:configurations}
\end{figure*}

Table~\ref{tab:posteriors} lists the median, $\pm1\sigma$, and $\pm2\sigma$ points in the posterior distributions obtained by fitting our models to only the NICER data, assuming a fully ionized hydrogen atmosphere, and the resulting maximum likelihood values for each of the parameters in these models. We display the complete corner plot of the posteriors from these same analyses in Figure~\ref{fig:posteriors}. Although the radius posterior is unimodal, many of the spot geometry posteriors are multimodal. We illustrate in Figure \ref{fig:configurations} the best-fitting spot configurations from three other modes, to complement that displayed in Figure \ref{fig:spots}.

\renewcommand{\arraystretch}{0.875}
\setlength{\tabcolsep}{2.pt}
\begin{deluxetable}{ccccccc}
\tablecaption{Three-circle fit to NICER Data on PSR~J0614$-$3329}
\tablehead{
\colhead{Parameter} & \colhead{median} & \colhead{$-1\sigma$} & \colhead{$+1\sigma$} & \colhead{$-2\sigma$} & \colhead{$+2\sigma$} & \colhead{Maximum Likelihood} 
}
\startdata
$R_{e}$ & 11.157 & 9.883 & 12.765 & 8.869 & 14.809 & 11.910 \\
$GM/c^2R_e$ & 0.190 & 0.166 & 0.214 & 0.144 & 0.236 & 0.185 \\
$M$ & 1.436 & 1.366 & 1.508 & 1.295 & 1.578 & 1.494 \\
$\theta_{\rm obs}$ & 1.541 & 1.537 & 1.545 & 1.533 & 1.550 & 1.542 \\
$\cos{\theta_{\rm c1}}$ & 0.017 & -0.484 & 0.511 & -0.917 & 0.920 & -0.057 \\
$\Delta \theta_1$ & 0.098 & 0.035 & 0.136 & 0.017 & 0.182 & 0.095 \\
$kT_{\rm eff,1}$ & 0.087 & 0.078 & 0.122 & 0.073 & 0.149 & 0.083 \\
$\cos{\theta_{\rm c2}}$ & -0.007 & -0.576 & 0.564 & -0.950 & 0.951 & -0.058 \\
$\Delta \theta_2$ & 0.099 & 0.033 & 0.155 & 0.016 & 0.383 & 0.021 \\
$kT_{\rm eff,2}$ & 0.085 & 0.060 & 0.125 & 0.038 & 0.150 & 0.132 \\
$\Delta \phi_2$ & 0.466 & 0.301 & 0.548 & 0.215 & 0.582 & 0.553 \\
$\cos{\theta_{\rm c3}}$ & -0.013 & -0.615 & 0.595 & -0.937 & 0.933 & 0.209 \\
$\Delta \theta_3$ & 0.123 & 0.034 & 1.629 & 0.016 & 2.427 & 0.098 \\
$kT_{\rm eff,3}$ & 0.063 & 0.035 & 0.120 & 0.026 & 0.146 & 0.062 \\
$\Delta \phi_3$ & 0.547 & 0.286 & 0.743 & 0.129 & 0.785 & 0.299 \\
$N_H$ & 0.597 & 0.167 & 1.441 & 0.023 & 2.639 & 0.034 \\
$d$ & 0.605 & 0.560 & 0.650 & 0.515 & 0.696 & 0.620 \\
$A_{\rm XTI}$ & 0.981 & 0.875 & 1.085 & 0.772 & 1.189 & 1.029 \\
\enddata
\tablecomments{A comparison of the $-2\sigma$, $-1\sigma$, median, $+1\sigma$, and $+2\sigma$, and maximum likelihood values inferred from our analysis of NICER data for PSR~J0614$-$3329.  Here we use our featured model, which has three uniform-temperature circular hot spots plus a phase-independent background.}
\label{tab:posteriors}
\end{deluxetable}

\bibliographystyle{aasjournal}
\bibliography{references}

\end{document}